\documentclass[]{aa}
\usepackage{psfig}

\begin{document}

\title{Detection of molecular hydrogen in a near Solar-metallicity damped
Lyman-$\alpha$ system at $z_{\rm abs}\approx 2$ toward
Q\,0551$-$366\thanks{Based on observations carried out at the European
Southern Observatory (ESO) under prog. ID No. 66.A-0624 with the
UVES spectrograph installed at the Very Large Telescope (VLT) on
Cerro Paranal, Chile.}}

\titlerunning{H$_2$ in a near Solar-metallicity DLA system at
$z_{\rm abs}\approx 2$}

\author{C. Ledoux\inst{1},
        R. Srianand\inst{2},
        P. Petitjean\inst{3,4}}

\offprints{C. Ledoux}

\institute{
   European Southern Observatory, Alonso de C\'ordova 3107, Casilla 19001, Vitacura, Santiago, Chile\\
   \email{cledoux@eso.org}
\and
   IUCAA, Post Bag 4, Ganesh Khind, Pune 411 007, India\\
   \email{anand@iucaa.ernet.in}
\and
   Institut d'Astrophysique de Paris -- CNRS, 98bis Boulevard Arago, 75014 Paris, France\\
   \email{petitjean@iap.fr}
\and
   LERMA, Observatoire de Paris, 61 Avenue de l'Observatoire, 75014, Paris, France
}

\date{Received 11 May 2002 / Accepted 24 June 2002}


\abstract{
We report the detection of H$_2$, C\,{\sc i}, C\,{\sc i}\,$^\star$,
C\,{\sc i}\,$^{\star\star }$ and Cl\,{\sc i} lines in a
near Solar-metallicity ([Zn/H$]=-0.13$) damped Lyman-$\alpha$ (DLA) system at
$z_{\rm abs}=1.962$ observed on the line of sight to the quasar Q\,0551$-$366.
The iron-peak elements, ${\rm X}={\rm Fe}$, Cr and Mn are depleted compared
to zinc, [X/Zn$]\sim -0.8$, probably because they are tied up onto
dust grains. Among the three detected H$_2$-bearing clouds, spanning 55
km s$^{-1}$ in velocity space, we derive a total molecular hydrogen
column density $N($H$_2)=2.6\times 10^{17}$ cm$^{-2}$ and a mean
molecular fraction $f=2N($H$_2)/(2N($H$_2)+N($H\,{\sc i}$))=1.7\times 10^{-3}$.
The depletion of heavy elements (S, Si, Mg, Mn, Cr, Fe, Ni and Ti) in the
central component is similar to that observed in the diffuse neutral gas
of the Galactic halo. This depletion is approximately the same in the
six C\,{\sc i}-detected components independently of the presence or absence of
H$_2$. The gas clouds in which H$_2$ is detected always have large
densities, $n_{\rm H}>30$ cm$^{-3}$, and low
temperatures, $T_{\rm 01}\la 100$ K. This shows that presence of dust,
high particle density and/or low temperature are required for molecules to
be present. The photo-dissociation rate derived in the components
where H$_2$ is detected suggests the existence of a local UV radiation field
similar in strength to the one in the Galaxy. Star formation therefore
probably occurs near these H$_2$-bearing clouds.
}

\maketitle

\section{Introduction}

Damped Lyman-$\alpha$ (DLA) absorption line systems observed in the spectrum of
quasars are associated with large H\,{\sc i} column
densities ($N$(H\,{\sc i}$)>2\times 10^{20}$ cm$^{-2}$). Therefore, they
are probably associated with regions of the Universe where star
formation occurs (Pettini et al. 1997). It is still unclear whether the
gas producing the absorption lines is located in
large, fast-rotating protogalactic
discs (Wolfe et al. 1986, Prochaska \& Wolfe 1997), in interacting building
blobs (Haehnelt et al. 1998) or in density fluctuations of
galaxy haloes (Ledoux et al. 1998).

Whatever the exact nature of DLA systems may be, molecules
(especially H$_2$) are expected to be found in those clouds. However,
despite intensive searches
(e.g. Black et al. 1987, Ge \& Bechtold 1999, Petitjean et al. 2000) only
four detections of H$_2$ have been reported to date. Recently, a fifth case
has been discovered serendipitously by Levshakov et al. (2002) at
$z_{\rm abs}=3.025$ toward Q\,0347$-$383.

The most recent survey for molecular hydrogen in 11 $z_{\rm abs}>1.8$
DLA systems, capitalizing on the unique capabilities of the UVES
high-resolution spectrograph of the ESO Very Large Telescope, has
given stringent upper limits, in the
range $1.2\times 10^{-7}$ -- $1.6\times 10^{-5}$, for the molecular
fraction $f=2N($H$_2)/(2N($H$_2)+N($H\,{\sc i}$))$ in nine of the systems
(Petitjean et al. 2000). Two possible detections have also been reported at
$z_{\rm abs}=2.374$ toward Q\,0841$+$129 (Petitjean et al. 2000) and at
$z_{\rm abs}=3.390$ toward Q\,0000$-$263 (Levshakov et al. 2000)
based, however, on the detection of only two weak features located into
the Lyman-$\alpha$ forest. Petitjean et al. (2000) concluded that
the non-detection of molecular hydrogen in most of the DLA systems could be
a direct consequence of high kinetic temperatures, $T>3000$ K, implying low
formation rates of H$_2$ onto dust grains. Therefore, most of the DLA systems
probably arise in warm and diffuse neutral gas. Fig.~5
of Petitjean et al. (2000) also suggests that, even if the gas is warm, H$_2$
should invariably be detected with $f\la 10^{-6}$. Such very low molecular
fractions probably result from high ambient UV flux.

In this paper, we present new results obtained from VLT-UVES high
resolution spectroscopy of a DLA system at $z_{\rm abs}=1.962$
toward Q\,0551$-$366. Details of the observations are given in Sect.~2.
The metal content, dust depletion pattern and H$_2$ molecular content of
the absorber are discussed in, respectively, Sects.~3, 4 and 5. Our
results are finally summarized and their implications discussed in Sect.~6.

\section{Observations}

The Ultraviolet and Visible Echelle Spectrograph (UVES; see
Dekker et al. 2000) mounted on the ESO Kueyen VLT-UT2 8.2 m telescope on
Cerro Paranal in Chile has been used in the course of a survey to search for
H$_2$ absorption lines in a sample of DLA systems.

High-resolution, high signal-to-noise ratio spectra of the
$m_{\rm V}=17.6$, $z_{\rm em}=2.32$ Q\,0551$-$366 quasar were obtained on
October 20-23, 2000. Standard settings were used in both blue and red arms
with Dichroic \#1 (central wavelengths: 3460 and 5800 \AA ) and central
wavelengths were adjusted to 4370 \AA\ in the Blue (7500 \AA\ in the Red)
with Dichroic \#2. This way, full wavelength coverage was obtained from
3050 to 7412 \AA\ and from 7566 to 9396 \AA\ accounting for the gap
between red-arm CCDs. The CCD pixels were binned $2\times 2$ in both arms and
the slit width was fixed to $1\arcsec$ yielding a spectral resolution
$R=42500$. The total integration times were 5 hours using Dichroic \#1 and
2h15min using Dichroic \#2.

The data have been reduced with the UVES
pipeline (Ballester et al. 2000) which is available as a context of the ESO
MIDAS data reduction system. The main characteristics of the pipeline is to
perform a precise inter-order background subtraction, especially for master
flat-fields, and to allow for an optimal extraction of the object
signal rejecting cosmic rays and performing sky-subtraction at the same time.
The pipeline products were checked step by step. We then converted the
wavelengths of the reduced spectra to vacuum-heliocentric values and
scaled, weighted and combined together individual 1D exposures using the NOAO
{\it onedspec} package of the IRAF software. The resulting unsmoothed spectra
were re-binned to the same wavelength step, 0.0471 \AA\ pix$^{-1}$,
yielding an average signal-to-noise ratio per resolution element as high
as 10 at 3200 \AA , 25 at 3700 \AA\ and 60 at 6000 \AA .

\section{Metal content}

\begin{figure}
\centerline{\vbox{
\psfig{figure=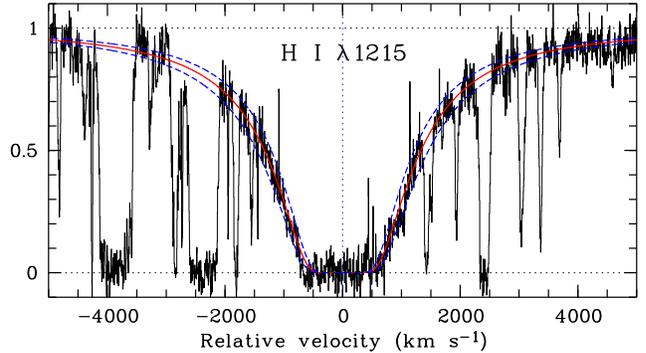,width=8.4cm,clip=,bbllx=53.pt,bblly=268.pt,bburx=542.pt,bbury=542.pt,angle=0.}}}
\caption[]{Portion of the normalized VLT-UVES spectrum showing the damped
Lyman-$\alpha$ line at $z_{\rm abs}=1.962$ toward Q\,0551$-$366. The best
Voigt-profile fitting, $N($H\,{\sc i}$)=(3.2\pm 0.6)\times 10^{20}$ cm$^{-2}$,
is superimposed onto the data.}
\label{figlya}
\end{figure}

\begin{figure}
\centerline{\vbox{
\psfig{figure=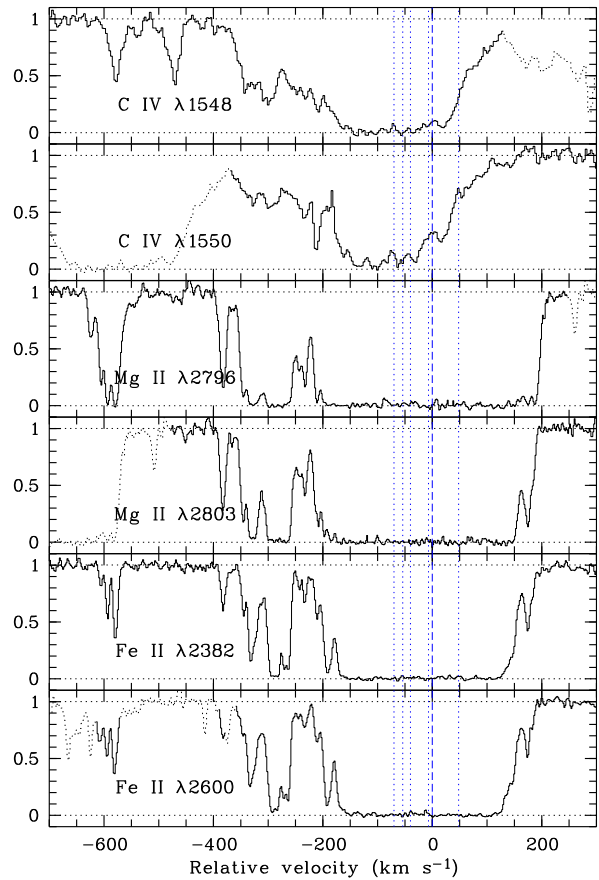,width=7.8cm,clip=,bbllx=53.pt,bblly=41.pt,bburx=541.pt,bbury=769.pt,angle=0.}}}
\caption[]{Absorption profiles related to the
C\,{\sc iv}\,$\lambda\lambda$1548,1550,
Mg\,{\sc ii}\,$\lambda\lambda$2796,2803 and
Fe\,{\sc ii}\,$\lambda\lambda$2382,2600 transition lines. Blended parts of the
profiles are shown by dotted histogrammes. The location of the
C\,{\sc i}-detected components (see Fig.~\ref{figcarbon}) is indicated by
vertical lines with the origin of the velocity scale corresponding
to $z_{\rm abs}=1.96221$.}
\label{figvelbro}
\end{figure}

\begin{figure*}
\centerline{\hbox{
\psfig{figure=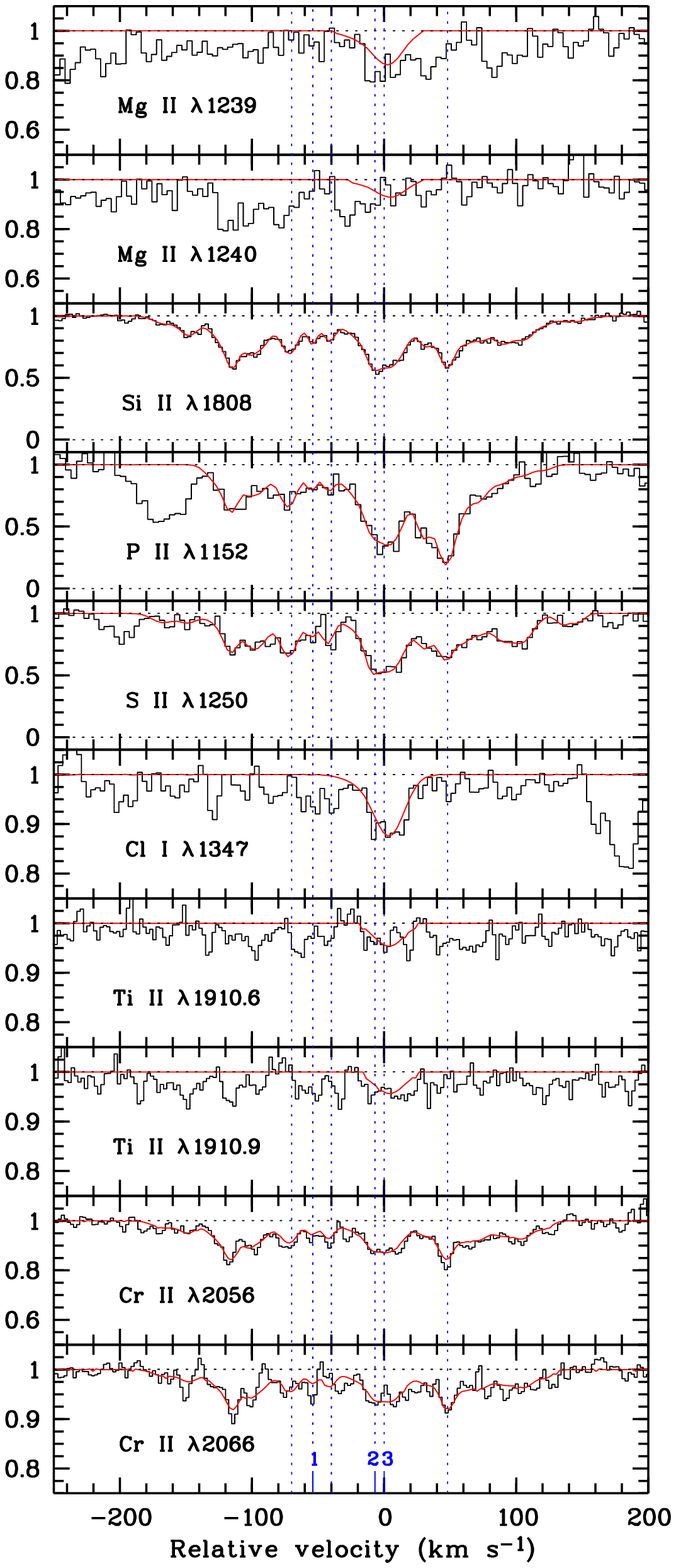,width=7.5cm,clip=,bbllx=60.pt,bblly=52.pt,bburx=369.pt,bbury=769.pt,angle=0.}\hspace{+0.3cm}\psfig{figure=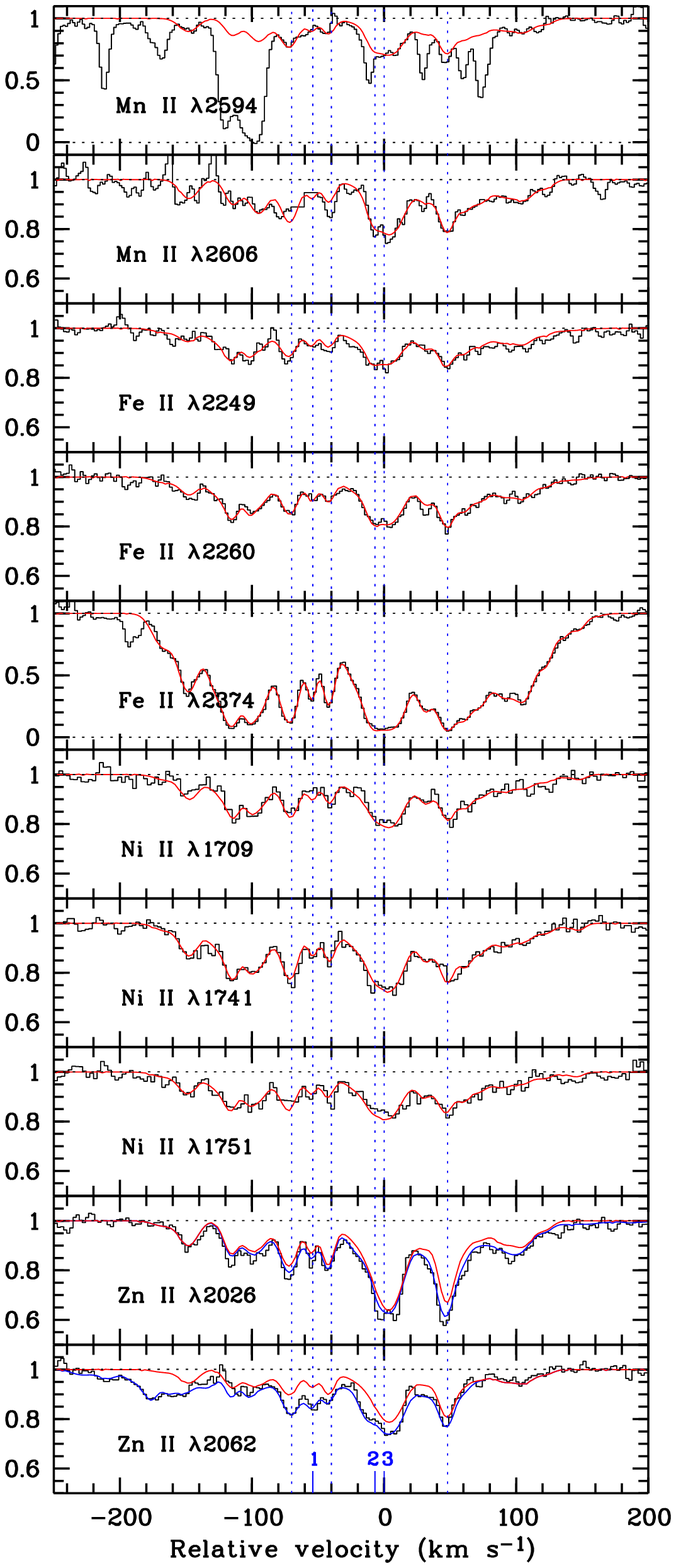,width=7.5cm,clip=,bbllx=60.pt,bblly=52.pt,bburx=369.pt,bbury=769.pt,angle=0.}}}
\vspace{0.3cm}
\centerline{\hbox{
\psfig{figure=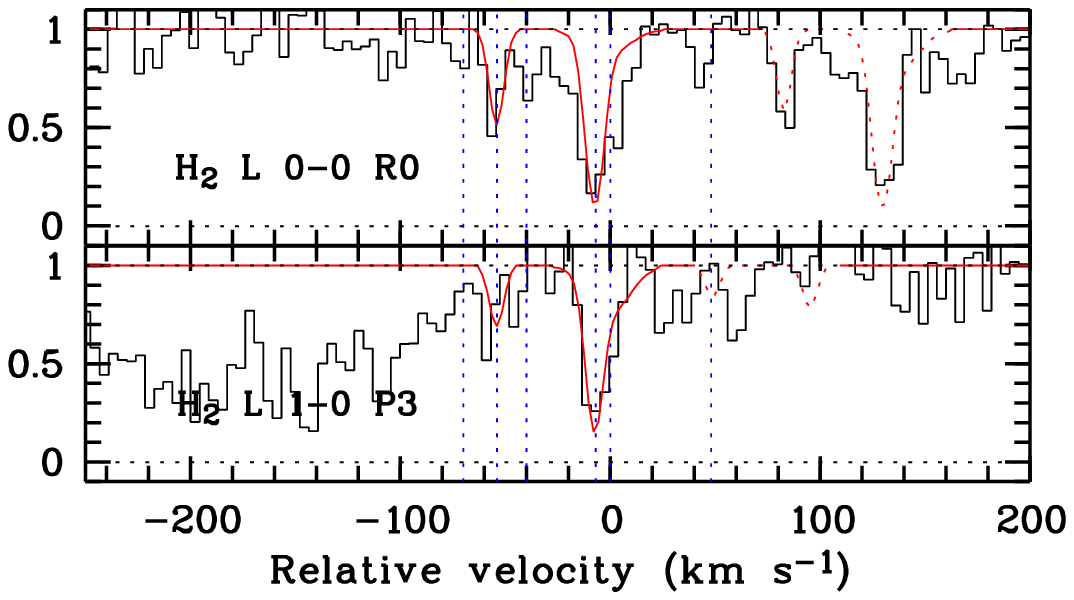,width=7.5cm,clip=,bbllx=60.pt,bblly=596.pt,bburx=369.pt,bbury=769.pt,angle=0.}\hspace{+0.3cm}\psfig{figure=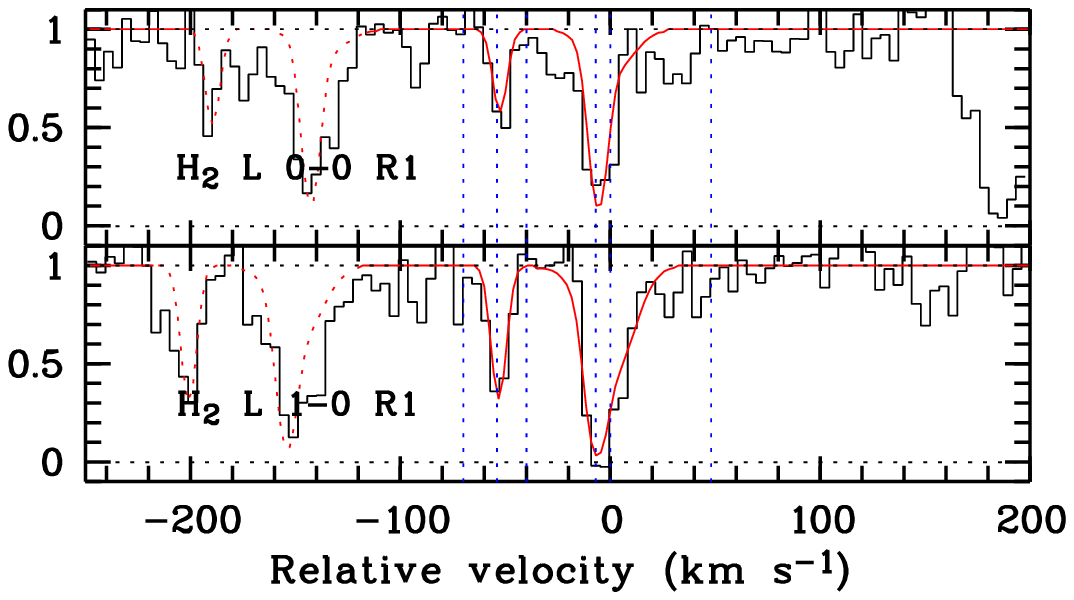,width=7.5cm,clip=,bbllx=60.pt,bblly=596.pt,bburx=369.pt,bbury=769.pt,angle=0.}}}
\caption[]{Comparison of the velocity profiles of metal lines
({\sl upper panels}) with a few of the H$_2$ lines ({\sl lower panels})
observed at $z_{\rm abs}=1.962$ toward Q\,0551$-$366. The best fitting is
superimposed onto the spectra. The position of the six components detected in
C\,{\sc i} is shown by vertical lines. The three components where H$_2$ is
detected are given numbers.\\ {\sl Upper panels}: for the fitting of
the Zn\,{\sc ii}\,$\lambda$2026 (resp. Zn\,{\sc ii}\,$\lambda$2062)
profile (light curves), the blending with Mg\,{\sc i}
(resp. Cr\,{\sc ii}) lines has been carefully taken into account by fitting
Mg\,{\sc i}\,$\lambda\lambda$2026,2852,
Cr\,{\sc ii}\,$\lambda\lambda\lambda$2056,2062,2066
and Zn\,{\sc ii}\,$\lambda\lambda$2026,2062 altogether (dark curves).
{\sl Lower panels}: dotted parts of the synthetic profiles correspond to
other H$_2$ transitions than the ones indicated.}
\label{figmetals}
\end{figure*}

\begin{table}
\caption{Atomic data}
\begin{tabular}{llll}
\hline
\hline
Transition                & $\lambda _{\rm vac}$ & $f$    & Ref.\\
                          & (\AA )               &        &     \\
\hline
H\,{\sc i}\,$\lambda$1215 & 1215.6701            & 0.4164 & a   \\
C\,{\sc i}\,$\lambda$1560                 & 1560.3092   & 0.0719   & b \\
C\,{\sc i}\,$^\star\lambda$1560.6         & 1560.6822   & 0.0539   & b \\
C\,{\sc i}\,$^\star\lambda$1560.7         & 1560.7090   & 0.0180   & b \\
C\,{\sc i}\,$^{\star\star }\lambda$1561.3 & 1561.3402   & 0.0108   & b \\
C\,{\sc i}\,$^{\star\star }\lambda$1561.4 & 1561.4384   & 0.0603   & b \\
C\,{\sc i}\,$\lambda$1656                 & 1656.9283   & 0.139    & b \\
C\,{\sc i}\,$^\star\lambda$1656.2         & 1656.2672   & 0.0589   & b \\
C\,{\sc i}\,$^\star\lambda$1657.3         & 1657.3792   & 0.0356   & b \\
C\,{\sc i}\,$^\star\lambda$1657.9         & 1657.9068   & 0.0473   & b \\
C\,{\sc i}\,$^{\star\star }\lambda$1657.0 & 1657.0082   & 0.104    & b \\
C\,{\sc i}\,$^{\star\star }\lambda$1658.1 & 1658.1212   & 0.0356   & b \\
Mg\,{\sc ii}\,$\lambda$1239   & 1239.9253 & 0.000554 & c \\
Mg\,{\sc ii}\,$\lambda$1240   & 1240.3947 & 0.000277 & c \\
Si\,{\sc ii}\,$\lambda$1808   & 1808.0129 & 0.00208  & d \\
P\,{\sc ii}\,$\lambda$1152    & 1152.8180 & 0.236    & a \\
P\,{\sc ii}\,$\lambda$1532    & 1532.5330 & 0.00761  & a \\
S\,{\sc ii}\,$\lambda$1250    & 1250.5780 & 0.00545  & a \\
Cl\,{\sc i}\,$\lambda$1188    & 1188.7742 & 0.0728   & a \\
Cl\,{\sc i}\,$\lambda$1347    & 1347.2396 & 0.153    & e \\
Ti\,{\sc ii}\,$\lambda$1910.6 & 1910.6090 & 0.104    & f \\
Ti\,{\sc ii}\,$\lambda$1910.9 & 1910.9380 & 0.098    & f \\
Cr\,{\sc ii}\,$\lambda$2056   & 2056.2569 & 0.105    & g \\
Cr\,{\sc ii}\,$\lambda$2062   & 2062.2361 & 0.0780   & g \\
Cr\,{\sc ii}\,$\lambda$2066   & 2066.1640 & 0.0515   & g \\
Mn\,{\sc ii}\,$\lambda$2594   & 2594.4990 & 0.271    & a \\
Mn\,{\sc ii}\,$\lambda$2606   & 2606.4619 & 0.193    & a \\
Fe\,{\sc ii}\,$\lambda$2249   & 2249.8768 & 0.00182  & h \\
Fe\,{\sc ii}\,$\lambda$2260   & 2260.7805 & 0.00244  & h \\
Fe\,{\sc ii}\,$\lambda$2374   & 2374.4603 & 0.0313   & i \\
Ni\,{\sc ii}\,$\lambda$1709   & 1709.6042 & 0.0324   & j \\
Ni\,{\sc ii}\,$\lambda$1741   & 1741.5531 & 0.0427   & j \\
Ni\,{\sc ii}\,$\lambda$1751   & 1751.9157 & 0.0277   & j \\
Zn\,{\sc ii}\,$\lambda$2026   & 2026.1371 & 0.489    & g \\
Zn\,{\sc ii}\,$\lambda$2062   & 2062.6604 & 0.256    & g \\
\hline
\hline
\end{tabular}
\label{tabosc}
\flushleft {\sc References:} Vacuum wavelengths from Morton (1991).\\ Oscillator strengths: (a)~Morton (1991); (b)~Wiese et al. (1996); (c)~Welty et al. (1999); (d)~Bergeson \& Lawler (1993b); (e)~Bi\'emont et al. (1994); (f)~Wiese et al. (2001); (g)~Bergeson \& Lawler (1993a); (h)~Bergeson et al. (1994); (i)~Bergeson et al. (1996); (j)~Fedchak et al. (2000).
\end{table}

From Voigt-profile fitting to the damped Lyman-$\alpha$ line, we find that the
total neutral hydrogen column density of the system at $z_{\rm abs}=1.962$
toward Q\,0551$-$366 is $\log N($H\,{\sc i}$)=20.50\pm 0.08$ (see
Fig.~\ref{figlya}). The associated metal line profiles of low-ionization
species are complex and span up to 850 km s$^{-1}$ in velocity space
as observed from the Mg\,{\sc ii}\,$\lambda\lambda$2796,2803
and Fe\,{\sc ii}\,$\lambda\lambda$2382,2600 transitions (Fig.~\ref{figvelbro}).
In Fig.~\ref{figmetals}, we display the strongest part of the profiles, in
the range [$-250$,$+200$] km s$^{-1}$ relative to $z_{\rm abs}=1.96221$, for a
large number of ions.
The (optically thin) line profiles were fitted together using the
{\it fitlyman} programme (Fontana \& Ballester 1995) running under
MIDAS, and/or {\it vpfit} (Carswell et al.,
http://www.ast.cam.ac.uk/{\small $\sim$}rfc/vpfit.html). The oscillator
strength values compiled in Table~\ref{tabosc} were used.

\begin{figure}
\centerline{\vbox{
\psfig{figure=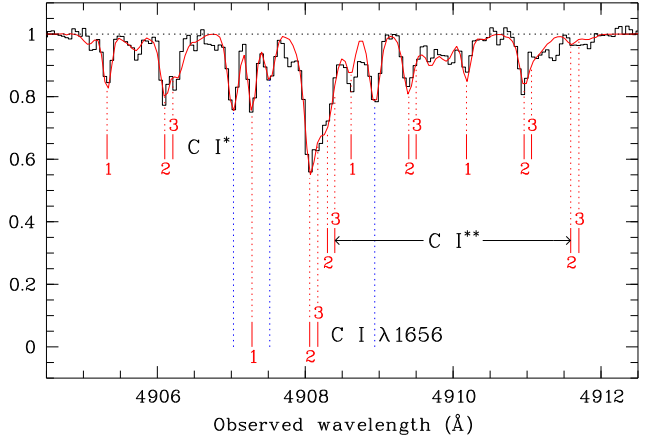,width=8.4cm,clip=,bbllx=60.pt,bblly=63.pt,bburx=552.pt,bbury=776.pt,angle=270.}}}
\caption[]{Absorption profile due to C\,{\sc i}\,$\lambda$1656.9,
C\,{\sc i}\,$^\star\lambda\lambda\lambda$1656.2,1657.3,1657.9 and
C\,{\sc i}\,$^{\star\star }\lambda\lambda$1657.0,1658.1 transition lines
at $z_{\rm abs}=1.962$. A total of six C\,{\sc i} components is detected over
120 km s$^{-1}$. The three components where H$_2$ and C\,{\sc i}\,$^\star$
absorption lines are both detected are indicated by numbers and
vertical bars.}
\label{figcarbon}
\end{figure}

\begin{figure*}
\centerline{\vbox{
\psfig{figure=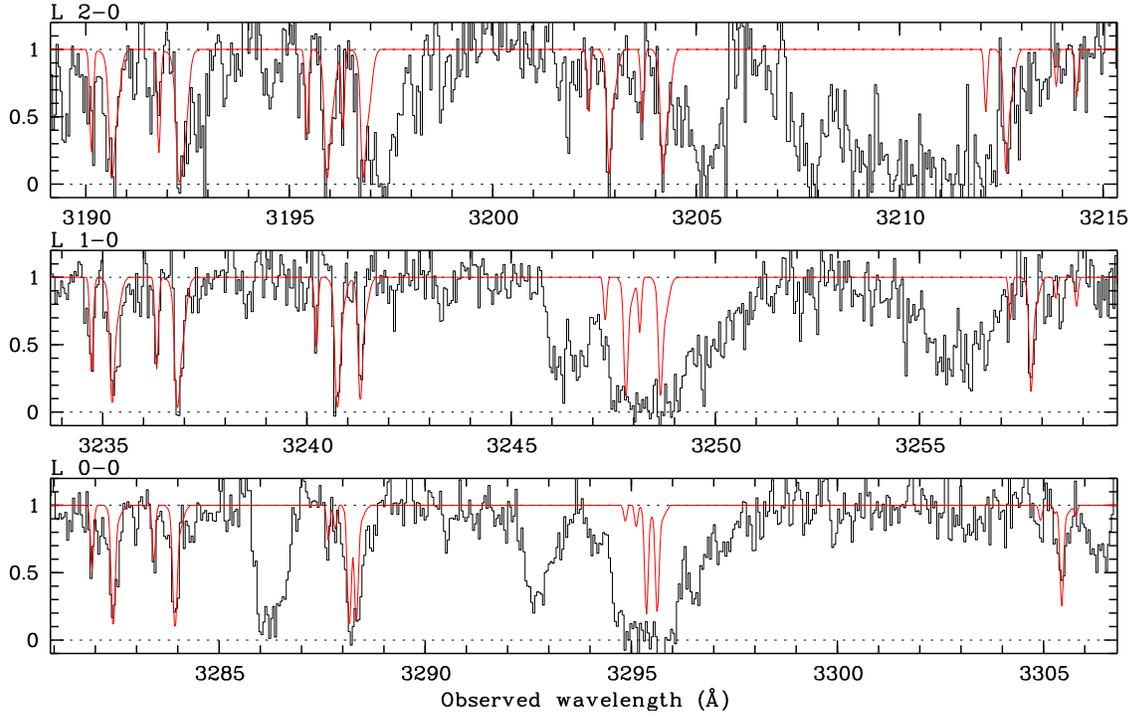,width=15.0cm,clip=,bbllx=102.pt,bblly=58.pt,bburx=563.pt,bbury=780.pt,angle=270.}}}
\caption[]{Selected transition lines from the $J=0$, 1, 2, 3 and 4
rotational levels of the vibrational ground-state of H$_2$
at $z_{\rm abs}=1.962$. H$_2$ molecules are detected in three different
clouds with velocity separations ranging from 7 to 55 km s$^{-1}$.}
\label{figmole}
\end{figure*}

Among the metal components, no less than six are detected in C\,{\sc i} whose
absorption line profiles are spread over $120$ km s$^{-1}$.
The C\,{\sc i}\,$\lambda\lambda$1560,1656 complexes have been
fitted simultaneously thereby ensuring uniqueness of the fitting
solution given that the relative positions of the C\,{\sc i},
C\,{\sc i}\,$^\star$ and C\,{\sc i}\,$^{\star\star }$ lines are different in
the two wavelength ranges. Among the C\,{\sc i} components, three
actually show absorption lines from C\,{\sc i}\,$^\star$, out of which two
also show lines from C\,{\sc i}\,$^{\star\star }$
(see Fig.~\ref{figcarbon}). Absorption lines due to different rotational
levels of H$_2$ in its vibrational ground state (see Fig.~\ref{figmole})
are detected in three of the components producing
detectable C\,{\sc i}\,$^\star$ lines.

\begin{table}
\caption{Metal abundances from the C\,{\sc i}-detected components of the
DLA system toward Q\,0551$-$366}
\begin{tabular}{clcc}
\hline
\hline
$z_{\rm abs}$ & Ion          & $\log N$        & [X/H$]^{\rm a}$\\
              &              &                 &                \\
\hline
1.962         & H\,{\sc i}   & $20.50\pm 0.08$ & ...            \\
              & Si\,{\sc ii} & $15.62\pm 0.06$ & $-0.43\pm 0.10$\\
              & P\,{\sc ii}  & $<14.03$        & $<-0.04$       \\
              & S\,{\sc ii}  & $15.38\pm 0.11$ & $-0.39\pm 0.14$\\
              & Cr\,{\sc ii} & $13.27\pm 0.06$ & $-0.91\pm 0.10$\\
              & Mn\,{\sc ii} & $13.11\pm 0.05$ & $-0.92\pm 0.09$\\
              & Fe\,{\sc ii} & $15.05\pm 0.05$ & $-0.96\pm 0.09$\\
              & Ni\,{\sc ii} & $14.08\pm 0.06$ & $-0.67\pm 0.10$\\
              & Zn\,{\sc ii} & $13.02\pm 0.05$ & $-0.13\pm 0.09$\\
\hline
\hline
\end{tabular}
\label{tabmet}
\flushleft $^{\rm a}$ Relative to Solar abundances (Savage \& Sembach 1996).
\end{table}


\begin{table}
\caption{Column densities of metal ions in C\,{\sc i}-detected components}
\begin{tabular}{llcc}
\hline
\hline
Ion & Transition & $\log N$    & $b$          \\
    & lines used &             & (km s$^{-1}$)\\
\hline
\multicolumn{4}{l}{$z_{\rm abs}=1.96150$}\\
Si\,{\sc ii} & 1808                     & $14.78\pm 0.03$\phantom{$^{\rm a}$}  & $8.4\pm 0.6$\phantom{$^{\rm b}$}\\
P\,{\sc ii}  & 1152,1532                & $12.99\pm 0.11$\phantom{$^{\rm a}$}  &   '' \\
S\,{\sc ii}  & 1250                     & $14.60\pm 0.10$\phantom{$^{\rm a}$}  &   '' \\
Cr\,{\sc ii} & 2056,2062,2066           & $12.44\pm 0.04$\phantom{$^{\rm a}$}  &   '' \\
Mn\,{\sc ii} & 2594,2606                & $12.37\pm 0.04$\phantom{$^{\rm a}$}  &   '' \\
Fe\,{\sc ii} & 2249,2260,2374$^{\rm a}$ & $14.27\pm 0.02$\phantom{$^{\rm a}$}  &   '' \\
Ni\,{\sc ii} & 1709,1741,1751           & $13.34\pm 0.03$\phantom{$^{\rm a}$}  &   '' \\
Zn\,{\sc ii} & 2026,2062                & $12.11\pm 0.05$\phantom{$^{\rm a}$}  &   '' \\
\hline
\multicolumn{4}{l}{$z_{\rm abs}=1.96167$}\\
Si\,{\sc ii} & 1808                     & $14.37\pm 0.05$\phantom{$^{\rm a}$}  & $3.1\pm 0.5$\phantom{$^{\rm b}$}\\
P\,{\sc ii}  & 1152,1532                & $12.48\pm 0.25$\phantom{$^{\rm a}$}  &   '' \\
S\,{\sc ii}  & 1250                     & $14.01\pm 0.27$\phantom{$^{\rm a}$}  &   '' \\
Cr\,{\sc ii} & 2056,2062,2066           & $11.94\pm 0.09$\phantom{$^{\rm a}$}  &   '' \\
Mn\,{\sc ii} & 2594,2606                & $11.70\pm 0.11$\phantom{$^{\rm a}$}  &   '' \\
Fe\,{\sc ii} & 2249,2260,2374$^{\rm a}$ & $13.78\pm 0.04$\phantom{$^{\rm a}$}  &   '' \\
Ni\,{\sc ii} & 1709,1741,1751           & $12.80\pm 0.06$\phantom{$^{\rm a}$}  &   '' \\
Zn\,{\sc ii} & 2026,2062                & $11.67\pm 0.06$\phantom{$^{\rm a}$}  &   '' \\
\hline
\multicolumn{4}{l}{$z_{\rm abs}=1.96180$}\\
Si\,{\sc ii} & 1808                     & $14.37\pm 0.08$\phantom{$^{\rm a}$}  & $4.9\pm 0.6$\phantom{$^{\rm b}$}\\
P\,{\sc ii}  & 1152,1532                & $12.51\pm 0.25$\phantom{$^{\rm a}$}  &   '' \\
S\,{\sc ii}  & 1250                     & $14.23\pm 0.19$\phantom{$^{\rm a}$}  &   '' \\
Cr\,{\sc ii} & 2056,2062,2066           & $12.14\pm 0.08$\phantom{$^{\rm a}$}  &   '' \\
Mn\,{\sc ii} & 2594,2606                & $11.91\pm 0.06$\phantom{$^{\rm a}$}  &   '' \\
Fe\,{\sc ii} & 2249,2260,2374$^{\rm a}$ & $13.89\pm 0.05$\phantom{$^{\rm a}$}  &   '' \\
Ni\,{\sc ii} & 1709,1741,1751           & $12.95\pm 0.06$\phantom{$^{\rm a}$}  &   '' \\
Zn\,{\sc ii} & 2026,2062                & $11.92\pm 0.05$\phantom{$^{\rm a}$}  &   '' \\
\hline
\multicolumn{4}{l}{$z_{\rm abs}=1.96203,1.96212,1.96225$$^{\rm b}$}\\
Mg\,{\sc ii} & 1239,1240                & $\le 15.42\pm 0.10$\phantom{$^{\rm a}$}  & $21.2\pm 1.1$$^{\rm b}$\\
Si\,{\sc ii} & 1808                     & $15.34\pm 0.06$\phantom{$^{\rm a}$}  &   '' \\
P\,{\sc ii}  & 1152,1532                & $<13.73$\phantom{$^{\rm a}$}         &   '' \\
S\,{\sc ii}  & 1250                     & $15.12\pm 0.07$\phantom{$^{\rm a}$}  &   '' \\
Cl\,{\sc i}  & 1188,1347                & $12.87\pm 0.09$\phantom{$^{\rm a}$}  &   '' \\
Ti\,{\sc ii} & 1910.6,1910.9            & $\le 12.40\pm 0.09$\phantom{$^{\rm a}$}  &   '' \\
Cr\,{\sc ii} & 2056,2062,2066           & $12.96\pm 0.06$\phantom{$^{\rm a}$}  &   '' \\
Mn\,{\sc ii} & 2594,2606                & $12.80\pm 0.04$\phantom{$^{\rm a}$}  &   '' \\
Fe\,{\sc ii} & 2249,2260,2374$^{\rm a}$ & $14.75\pm 0.05$\phantom{$^{\rm a}$}  &   '' \\
Ni\,{\sc ii} & 1709,1741,1751           & $13.79\pm 0.06$\phantom{$^{\rm a}$}  &   '' \\
Zn\,{\sc ii} & 2026,2062                & $12.74\pm 0.05$\phantom{$^{\rm a}$}  &   '' \\
\hline
\multicolumn{4}{l}{$z_{\rm abs}=1.96268$}\\
Si\,{\sc ii} & 1808                     & $14.94\pm 0.03$\phantom{$^{\rm a}$}  & $8.0\pm 0.6$\phantom{$^{\rm b}$}\\
P\,{\sc ii}  & 1152,1532                & $<13.58$\phantom{$^{\rm a}$}         &   '' \\
S\,{\sc ii}  & 1250                     & $14.63\pm 0.11$\phantom{$^{\rm a}$}  &   '' \\
Cr\,{\sc ii} & 2056,2062,2066           & $12.67\pm 0.03$\phantom{$^{\rm a}$}  &   '' \\
Mn\,{\sc ii} & 2594,2606                & $12.46\pm 0.05$\phantom{$^{\rm a}$}  &   '' \\
Fe\,{\sc ii} & 2249,2260,2374$^{\rm a}$ & $14.39\pm 0.03$\phantom{$^{\rm a}$}  &   '' \\
Ni\,{\sc ii} & 1709,1741,1751           & $13.35\pm 0.04$\phantom{$^{\rm a}$}  &   '' \\
Zn\,{\sc ii} & 2026,2062                & $12.39\pm 0.03$\phantom{$^{\rm a}$}  &   '' \\
\hline
\hline
\end{tabular}
\label{tabmet2}
\flushleft $^{\rm a}$ As well as 1608,2344,2382,2600.\\
$^{\rm b}$ The three components in the central part of the profile
are equivalent to first order to a single component at
$z_{\rm abs}=1.96221$ whose Doppler parameter is given.
\end{table}

\par\noindent
It is likely that most, if not all, of the neutral hydrogen originates from
the C\,{\sc i} components. Therefore, we derive total metal abundances
using column densities of singly ionized species summed up over the detected
C\,{\sc i} components (see Table~\ref{tabmet}). The result of fitting
to individual components is presented in Table~\ref{tabmet2}. Weak additional,
undetected C\,{\sc i} components cannot represent more than 15 per cent of
the total C\,{\sc i} column density (i.e. 0.06 dex). The best-fitting curves
shown in Fig.~\ref{figmetals} have been determined in two steps. In order
to derive accurate $b$ values, we first fitted transition lines
from Fe\,{\sc ii} covering a wide range of oscillator strengths with 22
components, six of which are C\,{\sc i}-detected. Using fixed $b$ values
as previously measured, we then fitted altogether the lines from Si\,{\sc ii},
P\,{\sc ii}, S\,{\sc ii}, Cr\,{\sc ii}, Mn\,{\sc ii}, Fe\,{\sc ii},
Ni\,{\sc ii} and Zn\,{\sc ii}. For a given component, the same Doppler
parameter and redshift were used for all species. Due to the smoothness of the
metal line profiles (see Fig.~\ref{figmetals}), we did not impose the Doppler
parameters from the fitting to C\,{\sc i} lines. It is well known that noise
leads to overestimate the width of weak lines. Small shifts ($<3$ km s$^{-1}$)
between the positions of the components in the metal and C\,{\sc i}
line profiles can be noted as well (see Tables~\ref{tabmet2}
and \ref{tabphy}). They are smaller than half of the FWHM of the observations
however. This implies that the uncertainty on the absolute metallicities is of
the order of 30\% but this is much smaller on the abundance ratios.

\par\noindent
The metallicity of the gas derived from the absolute abundance of zinc
is close to Solar, [Zn/H$]=-0.13$. This is the first time such a
high metallicity is observed in a DLA system at high redshift. The
abundance of S relative to Zn is not oversolar but is on the contrary
slightly undersolar, in agreement with the observations of Galactic thin
disk stars (see Chen et al. 2002 and refs. therein). The iron-peak elements
(${\rm X}={\rm Fe}$, Cr and Mn) are depleted compared to zinc,
[X/Zn$]\sim -0.8$, probably because they are tied up onto dust grains. This is
consistent with Si being slightly depleted compared to Zn.

\section{Dust depletion pattern}

\begin{figure}
\centerline{\vbox{
\psfig{figure=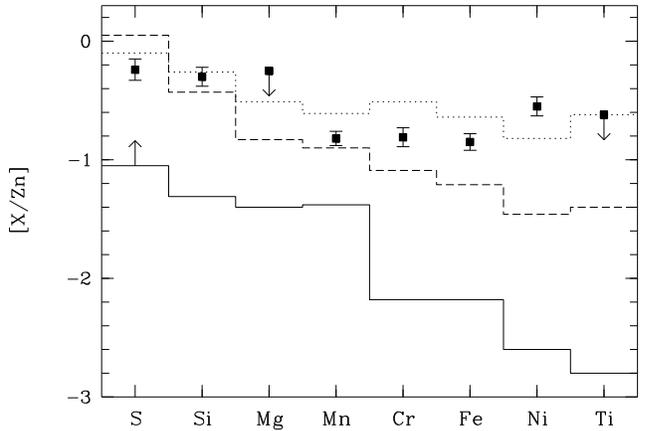,width=8.4cm,clip=,bbllx=51.pt,bblly=35.pt,bburx=555.pt,bbury=773.pt,angle=270.}}}
\caption[]{Depletion of heavy elements relative to zinc in the main sub-system
(components at $z_{\rm abs}=1.96203$, 1.96212 and 1.96225) of the
near Solar-metallicity DLA system toward Q\,0551$-$366. The histogrammes
show the values observed in cold (continuous line) and warm (dashed line)
Galactic disc clouds, and warm Galactic halo clouds (dotted line;
from Sembach \& Savage 1996).}
\label{figdeple}
\end{figure}

We compare in Fig.~\ref{figdeple} the abundance pattern observed in the main
(central) component, where absorption lines from a large number of
heavy elements are detected (see Table~\ref{tabmet2}), with the abundance
patterns observed in different Galactic environments. To compute the depletion
of an element relative to the abundance of zinc, we use the
definition [X/Zn$]=\log [N($X$)/N($Zn$)]-\log [N($X$)/N($Zn$)]_{\odot}$
together with the Solar abundances from Savage \& Sembach (1996). The error
bars on the measurements are $1\sigma$ uncertainties. The continuous
and dashed lines are the mean observed trends in the Galactic disc for cold
and warm gas respectively, while the dotted line shows the abundance pattern
of diffuse gas in the Galactic halo (from Sembach \& Savage 1996). It is
clear from Fig.~\ref{figdeple} that the relative abundance pattern of
this near Solar-metallicity DLA system is close to that observed in
Galactic warm halo clouds.

\begin{figure}
\centerline{\vbox{
\psfig{figure=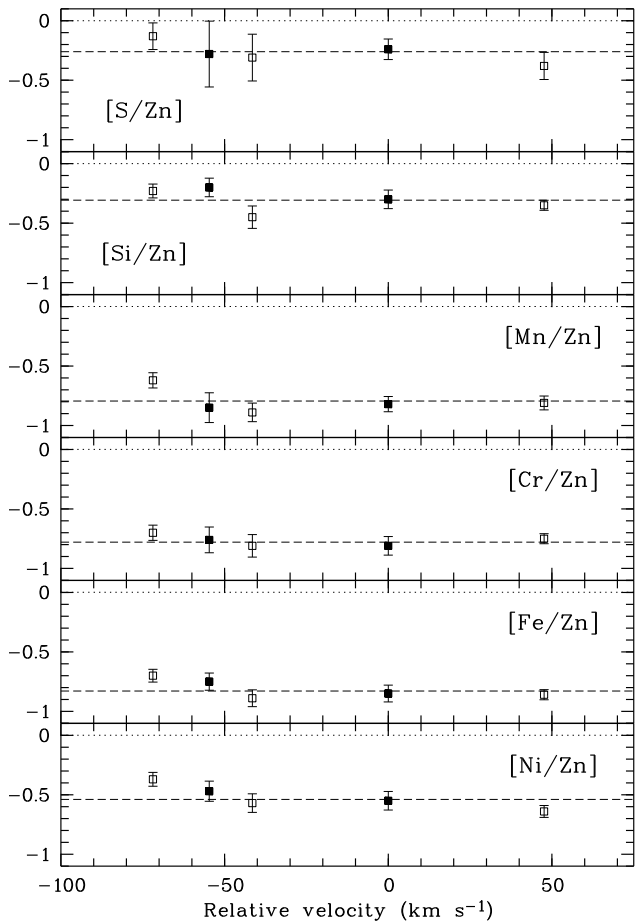,width=8.4cm,clip=,bbllx=40.pt,bblly=40.pt,bburx=542.pt,bbury=769.pt,angle=0.}}}
\caption[]{Depletion of heavy elements relative to Zn in the velocity
sub-systems where C\,{\sc i} is detected. The metal components
at $z_{\rm abs}=1.96203$, 1.96212 and 1.96225 (see Table~\ref{tabmet2})
are considered as a single sub-system at $z_{\rm abs}=1.96221$ ($V=0$ km
s$^{-1}$). The horizontal dashed lines show the column density weighted-mean
values of the ratios. Although the depletion pattern is similar in every
components, only two of them (shown as filled squares) have
detectable H$_2$ lines: {\it this demonstrates that the presence of H$_2$
is not only related to the dust-to-metal ratio}.}
\label{figratios}
\end{figure}

So as to probe any possible velocity dependence of the depletion pattern,
we show in Fig.~\ref{figratios} the metal abundance ratios [X/Zn] measured
in the C\,{\sc i}-detected components. The horizontal dashed lines give the
column density weighted-mean values of the abundance ratios. It is apparent
that every component, whether absorption lines due to H$_2$ are present
or not, show similar abundance patterns, and as a consequence
similar dust-to-metal ratios, within observational uncertainties. This clearly
demonstrates that in this DLA system the presence of molecules is not
only related to the dust-to-metal ratio.
This also suggests in this case that the gas has undergone uniform mixing of
heavy elements.

A word of caution should be said here. Whereas the $-55$ km s$^{-1}$
component is clearly identified in the metal line profiles, this is not
the case of the central components. The overall profiles of most of the metal
lines are smooth with no predominant component. However, it is apparent from
the Si\,{\sc ii} profile that the narrow component at $z_{\rm abs}=1.96214$
($V=-7$ km s$^{-1}$) is present at the bottom of the profile. {\it This
means that absorption lines due to the cold gas, i.e. related
to C\,{\sc i} components, can possibly be hidden by a more pervasive
medium}. This implies that absolute metallicities are not
accurately determined in the neutral gas phase. This is probably not important
for volatile elements as the dominant components actually produce the
strongest lines, but could be of importance for Fe co-production elements
which can be depleted. Consequently, absorption lines from the latter
elements can be weak and lost in the overall profiles, which means we may
underestimate the depletion factors in the central components and, in
particular, the one at $z_{\rm abs}=1.96214$. As an example, a large
depletion factor has been derived in a well-defined, weak H$_2$
component at $z_{\rm abs}=1.968$ observed on the line of sight to
Q\,0013$-$004 (Petitjean et al. 2002).

Note that Cl\,{\sc i} is detected
in the main component of the DLA system toward Q\,0551$-$366, at
$z_{\rm abs}=1.96221$, with $\log N($Cl\,{\sc i}$)=12.87\pm 0.09$ (see
Table~\ref{tabmet2}). This translates to [Cl/Zn$]=-0.49\pm 0.10$ under the
assumption that [Cl\,{\sc i}$]\equiv [$Cl]. This is close to what is observed
in the Galactic interstellar medium (ISM) toward 23\,Ori (Welty et al. 1999).
The small value of this ratio is probably a consequence of Cl\,{\sc i} being
partially ionized. This first detection of neutral chlorine in a DLA system
should be related to the high metallicity of the gas as well as to
the presence of H$_2$ with which Cl\,{\sc ii} rapidly reacts to form
Cl\,{\sc i}.

\section{H$_2$ molecular content}

\begin{table}
\caption{Voigt-profile fitting results for different rotational levels of the
vibrational ground-state transition lines of H$_2$}
\begin{tabular}{cccccc}
\hline
\hline
$z_{\rm abs}$ & Level & $\log N($H$_2)$ & $b$           & $T_{\rm ex}$ &J$_{\rm ex}$\\
              &       &                 & (km s$^{-1}$) & (K)          &            \\
\hline
1.96168 & $J=0$ &  $15.19^{+0.46}_{-0.13}$ &  $2.1\pm 0.7$ &     ...       &   \\
        & $J=1$ &  $15.23^{+0.33}_{-0.14}$ &   ''          &  $76 \pm   7$ &0-1\\
        & $J=2$ &  $14.76^{+0.27}_{-0.09}$ &   ''          &  $248\pm  52$ &0-2\\
        & $J=3$ &  $14.72^{+0.18}_{-0.06}$ &   ''          &  $281\pm  36$ &0-3\\
        & $J=4$ & $<14.19^{\rm a}$         &   ''          & $<680$        &0-4\\
        & $J=5$ & $<14.34^{\rm a}$         &   ''          & $<671$        &0-5\\
1.96214 & $J=0$ &  $16.83^{+0.47}_{-1.09}$ &  $2.1\pm 0.8$ &     ...       &   \\
        & $J=1$ &  $17.12^{+0.46}_{-0.69}$ &   ''          &  $81 \pm  29$ &0-1\\
        & $J=2$ &  $16.56^{+0.84}_{-0.80}$ &   ''          &  $378\pm 255$ &0-2\\
        & $J=3$ &  $16.24^{+1.12}_{-0.59}$ &   ''          &  $378\pm 243$ &0-3\\
        & $J=4$ &  $14.35^{+0.05}_{-0.13}$ &   ''          &  $420\pm 178$ &0-4\\
        & $J=5$ & $<14.34^{\rm a}$         &   ''          & $<556$        &0-5\\
1.96221 & $J=0$ &  $14.74^{+0.05}_{-0.13}$ & $12.8\pm 1.7$ &     ...           \\
        & $J=1$ &  $15.18^{+0.03}_{-0.05}$ &   ''          &  $97 \pm  40$ &0-1\\
        & $J=2$ &  $14.93^{+0.02}_{-0.16}$ &   ''          &  $357\pm  69$ &0-2\\
        & $J=3$ &  $14.96^{+0.01}_{-0.11}$ &   ''          &  $352\pm  43$ &0-3\\
\hline
\hline
\end{tabular}
\label{tabmol}
\flushleft $^{\rm a}$ $3\sigma$ upper limit.
\end{table}

Lyman-band absorption lines from H$_2$ are detected in
two well-detached sub-systems at $z_{\rm abs}=1.96168$ and
1.9622 ($\Delta V\approx 55$ km s$^{-1}$) (see Fig.~\ref{figmetals}
and \ref{figmole}). The former sub-system, corresponding to
the C\,{\sc i} component labelled ``1'' in Fig.~\ref{figmetals}, is narrow and
weak but clearly detected in the $J=0$, 1, 2 and 3 rotational levels of the
vibrational ground-state of H$_2$. The latter sub-system, corresponding to
C\,{\sc i} components ``2'' and ``3'', is broader and stronger. In this
case, all unblended lines from the $J=0$, 1, 2 and 3 levels have been fitted
together assuming the presence of two components at $z_{\rm abs}=1.96214$ and
1.96221 as observed in C\,{\sc i}. The column density for each $J$ level was
determined from several trials of Voigt-profile fitting using the
oscillator strength values from Morton \& Dinerstein (1976). The results
are given in Table~\ref{tabmol}. Errors in the column densities were estimated
taking into account the range of possible $b$ values (from the fitting
to C\,{\sc i} lines) and the level of saturation introduced by background
subtraction uncertainties ($\sim 5$\%). Errors in the column
densities therefore correspond to a {\it range of column densities} and not to
the rms error from fitting the Voigt profiles. The fifth Column of
Table~\ref{tabmol} gives the measured excitation temperatures relative to the
$J=0$ level (see Srianand \& Petitjean 1998). The mean and the error in
the excitation temperatures were obtained from all the
individual Voigt-profile fittings previously discussed.

The $z_{\rm abs}=1.96214$ component alone corresponds to 95 per cent of
the total H$_2$ column density of the absorber, $N($H$_2)=2.6\times 10^{17}$
cm$^{-2}$. Note that, even if $b<1$ km s$^{-1}$ in the $z_{\rm abs}=1.96214$
component, the total H$_2$ column density can not exceed $10^{18}$ cm$^{-2}$.
Unfortunately, the molecular fractions in individual components cannot
be determined as the respective H\,{\sc i} column densities are not known.
The mean molecular fraction along the line of sight is $f=1.7\times 10^{-3}$.
The kinetic temperature in H$_2$-bearing clouds can be approximated to a first
order by the excitation temperature $T_{01}$ measured between the $J=0$ and
1 levels. We find $T_{01}=76\pm 7$, $81\pm 29$ and $97\pm 40$ K
at $z_{\rm abs}=1.96168$, 1.96214 and 1.96221 respectively. These values
are typical of what is measured in the halo of our Galaxy and along lines of
sight through the LMC and SMC (Shull et al. 2000, Tumlinson et al. 2002)
and are similar to what is derived in different components
at $z_{\rm abs}=2.34$ toward Q\,1232$+$082 and $z_{\rm abs}=1.97$ toward
Q\,0013$-$004 (Srianand et al. 2000, Petitjean et al. 2002).

Excitation temperatures for higher $J$ levels are also given in
Table~\ref{tabmol}. In the case of the components at, respectively,
$z_{\rm abs}=1.96168$ and 1.96221 toward Q\,0551$-$366, the high $J$ level
populations can be explained by a single excitation temperature within
measurement uncertainties (respectively $T_{\rm ex}\approx 270$ and 350
K). These values are larger than the measured kinetic temperatures.
This suggests that, in addition to collisions, other processes like UV pumping
and formation pumping are at play to populate these levels.
In the case of the $z_{\rm abs}=1.96214$ component, line
blending, uncertainty in the position of the zero level and
possible saturation of the low $J$ lines actually result in large errors
in the column density measurement, and a single excitation temperature,
$T_{\rm ex}\approx 380$ K, is also consistent with the observed high $J$
level populations. Incidentally, similar population ratios are observed
along lines of sight through the Magellanic stream
(Sembach et al. 2001, Richter et al. 2001).

\section{Discussion}

\begin{table*}
\caption{Physical conditions in individual components}
\begin{tabular}{cccccccc}
\hline
\hline
$z_{\rm abs}$ & $\log N($C\,{\sc i}$)$ & $\log N($C\,{\sc i}\,$^\star)$ &
$\log N($C\,{\sc i}\,$^{\star\star })$ & $b$           &
$T_{\rm CMB}$ & $T_{\rm 01}$     & $n_{\rm H}$\\
              &                        &                                &
                                       & (km s$^{-1}$) &
(K)           & (K)              & (cm$^{-3}$)\\
\hline
1.96152 & $12.69\pm 0.07$  & $<12.18$\phantom{\,:} & $<11.94$\phantom{\,:} & \phantom{0}$4.3\pm 1.4$ & $\le 10.5$ &   ...       & $<17$    \\
1.96168 & $12.64\pm 0.07$  & $12.84\pm 0.07$       & $<12.16$\,:           & \phantom{0}$2.1\pm 0.7$ & $\le 22.3$ & $76\pm \phantom{0}7$   & $170-185$\\
1.96180 & $12.42\pm 0.13$  & $<12.18$\phantom{\,:} & $<11.94$\phantom{\,:} & \phantom{0}$3.9\pm 2.3$ & $\le 12.9$ &   ...       & $<56$    \\
1.96214 & $12.66\pm 0.12$  & $12.69\pm 0.11$       & $12.11\pm 0.34$       & \phantom{0}$2.1\pm 0.8$ & $\le 18.7$ & $81\pm 29$  & $\phantom{0}55-390$ \\
1.96221 & $13.16\pm 0.06$  & $12.98\pm 0.09$       & $12.26\pm 0.36$       &           $12.8\pm 1.7$ & $\le 14.5$ & $97\pm 40$  & $\phantom{0}30-150$ \\
1.96268 & $12.63\pm 0.08$  & $<12.18$\phantom{\,:} & $<11.94$\phantom{\,:} & \phantom{0}$4.0\pm 1.8$ & $\le 10.0$ &   ...       & $<25$    \\
\hline
\hline
\end{tabular}
\label{tabphy}
\end{table*}


Considering the cosmic microwave background radiation (CMBR) to be the
only source of C\,{\sc i} excitation (e.g. Srianand et al. 2000), we can
derive upper limits on the CMBR temperature from the column densities of
C\,{\sc i}, C\,{\sc i}\,$^\star$ and C\,{\sc i}\,$^{\star\star }$ that we
measure in different components of the DLA system toward Q\,0551$-$366 (see
Table~\ref{tabphy}). In the components where H$_2$ is detected, the values
are found to be much larger than what is predicted from standard Big-Bang
cosmology (i.e. about 8 K). Under the conditions prevailing in these
clouds, fluorescence is negligible in populating the excited levels of
C\,{\sc i} (see Fig.~2 of Silva \& Viegas 2002). This means that excitation
by collisions is important and, therefore, that the influence of
local physical conditions (both density and temperature) is substantial in
those clouds. Conversely, assuming the CMBR temperature to be 8 K and
the kinetic temperature being approximated by the excitation temperature of
the $J=1$ rotational level, we can estimate the hydrogen density from
the relative populations of the C\,{\sc i} fine-structure levels. For the
components where H$_2$ is not detected, we assume $T_{\rm kin}=100$ K. As
can be seen from Table~\ref{tabphy}, it is apparent that the densities
are larger in the gas where H$_2$ is detected.
Therefore, the non-detection of H$_2$ in components having
{\it similar dust-to-metal ratios and similar column densities} of
heavy elements is a consequence of lower densities and possibly
higher temperatures.
This is in line with the conclusions of Petitjean et al. (2000). The allowed
range in pressure $p/k$ measured in the three H$_2$ components
are respectively 12950$-$13770, 6050$-$20280 and 4170$-$8550 cm$^{-3}$ K.
Such high pressures are seen only in $\le 3$\% of the C\,{\sc i} gas of our
Galaxy (Jenkins \& Tripp 2001) but are consistent with the high
pressures also derived in the case
of Q\,0013$-$004 (Petitjean et al. 2002). They probably indicate that the
gas under consideration is undergoing high compression due to either
collapse, merging and/or supernovae explosions.


It is difficult to measure the rate of H$_2$ formation onto dust grains in DLA
systems because it is related to the amount of dust which is itself difficult
to estimate. In the present DLA system, even if the mean metallicity is close
to solar the observed mean depletion of Fe is an order of magnitude smaller
than what is typically measured in Galactic cold disc clouds. Therefore, we
can estimate that the rate of H$_2$ formation onto dust grains is about
one tenth of the Galactic value, thus $R\sim 3\times 10^{-18}$ cm$^3$
s$^{-1}$. Using the simple model proposed by Jura (1975) and the
column densities and densities estimated above, we derive
the photo-absorption rate of the Lyman and Werner bands
of H$_2$: $\beta_0=1.4\times 10^{-11}-2.8\times 10^{-10}$ s$^{-1}$ for
the component at $z_{\rm abs}=1.96214$.
This value is similar to
the average value measured in the Galactic
ISM, $\beta_0=5\times 10^{-10}$ s$^{-1}$, while the average for the
intergalactic radiation field is $\beta_0\approx 2\times 10^{-12}$ s$^{-1}$
with $J_{21}({\rm 912~\AA })=1$. This shows that there
is in-situ star-formation close to the absorbing clouds and that H$_2$
can survive, at least with such molecular fractions, even in the presence of a
strong UV radiation field because the gas has a high particle density.
One should keep in mind that the rate of H$_2$ formation goes linearly
with the density of dust grains while it goes as the second power of
the H\,{\sc i} density. Therefore, it is natural that, even though the
presence of dust is a crucial factor for the formation of molecules,
{\it local physical conditions} such as particle density, temperature
and local UV field, {\it play an important role in governing the molecular
fraction of a given cloud} in DLA systems.

\begin{acknowledgements}
It is a pleasure to thank R. Carswell and A. Smette for their assistance
in using the latest versions of the {\it vpfit} and {\it fitlyman}
programmes respectively. CL acknowledges support from an
ESO post-doctoral fellowship. PP thanks ESO, and in particular D. Alloin,
for an invitation to stay at the ESO headquarters in Chile where part of
this work was completed. This work was supported by the European Community
Research and Training Network ''The Physics of the Intergalactic Medium''. RS
and PP gratefully acknowledge support from the Indo-French Centre for
the Promotion of Advanced Research (Centre Franco-Indien pour la Promotion
de la Recherche Avanc\'ee) under contract No. 1710-1.
\end{acknowledgements}

\end{document}